\documentclass[runningheads]{llncs}
\usepackage[T1]{fontenc}
\usepackage{graphicx,verbatim}
\usepackage{amsfonts}
\usepackage{amsmath}
\usepackage{amssymb}
\usepackage{esvect}
\usepackage{siunitx}
\usepackage{multirow}
\usepackage{booktabs}
\usepackage{multirow}
\usepackage{graphicx}  
\begin{document}
\title{Unsupervised Adversarial Domain Adaptation for Uterine layer Segmentation: From Labeled Cine to Unlabeled Dynamic EPI MRI}
\titlerunning{Peristalsis}
 \author{Smiti Tripathy\inst{1} \and
 Milauni Desai\inst{1}\and
 Jordina Aviles Verdera\inst{2,3}\and
 Jana Hutter\inst{2,3}}

\authorrunning{S Tripathy et al.}

\institute{Institute of Radiology, University Hospital Erlangen, Friedrich-Alexander-Universität Erlangen-Nürnberg, Erlangen, Germany \and CAIMED, L3S, Hannover, Germany\and  Institut für Informationsverarbeitung, Leibniz University Hannover, Germany}
\maketitle              
\begin{abstract}
Uterine peristalsis is a key physiological phenomenon responsible for various functions across the menstrual cycle, intimately linked to uterine wall microstructure. Alterations in uterine motion and tissue properties are implicated in the etiology of gynecological diseases, yet these processes have been studied in isolation. We introduce a dynamic multi-echo gradient echo EPI framework for simultaneous characterization and correlation of uterine peristaltic activity and time-resolved T2* changes at 0.55T. Inherent susceptibility artifacts, reduced resolution, and burden of manual uterine layer annotation are addressed by an unsupervised adversarial domain adaptation framework, transferring segmentation knowledge from labeled 
cine MRI to unlabeled dynamic EPI. We implemented Unet-LSTM with multi-scale domain discriminators that exploits temporal layer dynamics. A Dice score of 0.88 and Jaccard index of 0.80 was achieved. Mean T2* values were 108ms, 76ms, and 124ms for the myometrium, junctional zone, and endometrium. A negative correlation between junctional zone area and T2* was observed in 14/39 cases, providing first insights into oxygenation patterns associated with junctional zone contraction and motion, demonstrating feasibility of assessing the interplay between contractility and dynamic T2* changes.
 

\end{abstract}

\section{Introduction}
\noindent The intrinsic motion of the uterus, peristalsis, is a fascinating physiological phenomenon, varying with hormonal state. It plays a key role in reproductive function by facilitating sperm transport for fertilization and the expulsion of menstrual blood \cite{kunz1996dynamics,de1990contractions}. Deviations in the carefully coordinated motion patterns are implicated in the etiology of adenomyosis and endometriosis \cite{Shaked2014} and play a role in infertility \cite{Kuijsters2017}. Uterine peristalsis is enabled by the interwoven muscular layers in the uterine wall. The microstructure (e.g. vasculature, collagen, and fibers) and motion are therefore closely intertwined. Understanding and quantifying both the tissue properties in the uterine wall and peristaltic motion carries great potential for research and clinical applications. The excellent soft-tissue contrast of MRI, combined with its capability for dynamic in-vivo cine imaging, enables visualization and quantitative assessment of uterine peristalsis \cite{Nakai2012,Celli2022}. Furthermore, T2* relaxometry, based on the blood oxygenation level-dependent effect, has facilitated the evaluation of tissue oxygenation patterns, providing functional insights into uterine physiology \cite{kido2007physiological,li2022cyclic}. However, these approaches till date are limited to motion or microstructure in isolation.

Deep learning–based segmentation networks are highly effective when trained on large amounts of labeled data; however, their performance degrades when applied to unseen domains. Several studies have demonstrated the effectiveness of adversarial domain adaptation techniques, transferring knowledge from labeled high-resolution source domains to unlabeled or low-resolution target domains \cite{zheng2025unsupervised,xie2022unsupervised,orbes2019multi}. Existing uterine segmentation models focus predominantly on high-resolution anatomical MRI \cite{cui2024segmentation,kurata2019automatic,angeline2025mri}, there is currently no established segmentation pipeline for dynamic or functional uterine MRI. The limited availability of annotated datasets, combined with the dynamic changes in the uterine-layered structure and variable susceptibility effects, further constrain the transfer of supervised models to different contrasts. Non-quantitative dynamic data is often acquired with balanced steady-state free precession sequences, providing high-contrast visualization of uterine anatomy, whereas echo-planar imaging (EPI)-based functional sequences suffer from reduced spatial resolution and increased vulnerability to susceptibility artifacts. These differences introduce significant distribution shifts. This phenomenon, commonly referred to as domain shift, poses a major challenge for robust learning, as models trained on a labeled source domain often fail to generalize effectively to the target domain.

Main contributions of this work are: (1) intertwined assessment of uterine motion and layer-specific tissue composition using 0.55T MRI, (2) a U-Net–LSTM-based segmentation framework with multi-level adversarial domain discriminators for unsupervised domain adaptation from labeled cine to unlabeled dynamic EPI data, (3) demonstration of successful simultaneous assessment of interlinked junctional zone dynamics and T2* variations.

\section{Methods}
\subsection{Data acquisition and MRI Scan Protocol}
\noindent In this prospective study, uterine MRI was performed on a clinical 0.55T Magnetom Free.Max scanner (Siemens Healthineers) using a 9-channel contour-M coil and 6-channel spine coil in supine position in 100 female participants after written informed consent. No bowel preparation was performed and no antispasmodics were administered. After anatomical imaging, a sagittal cine MRI sequence (FOV=$290\times290 mm$, 5 slices, $1.13\times1.13\times6 mm$, TR= 335.1-1000 ms, TE=2.09ms, 80-250 frames, time= 7-10 minutes) was acquired along the uterine cavity. Immediately after cine imaging, sagittal dynamic MEGE EPI data was acquired (FOV = $294\times294$, 8 slices, $3.062\times3.062\times6 mm$, GRAPPA 2, TR=1700 ms, TE= [31 ms, 79ms, 127ms], 250 dynamics,time = 7 minutes). No manual shimming was performed. Both sequences were acquired free-breathing.

\subsection{Domain Adaptation Segmentation Network}
The proposed domain adaptation segmentation network comprises two branches: a main segmentation branch (Unet-LSTM based) and an adversarial domain discriminator branch. The segmentation backbone follows a U-net architecture with three encoder stages, a bottleneck, and three decoder stages. Convolutional long short-term memory (C-LSTM) blocks were incorporated followed by a single 2D convolutional layer in each encoder and bottleneck stages to capture temporal dependencies. Decoder stages consisted of double 2D convolutional layers. Each convolutional layer was followed by a LeakyReLU and 2D instance normalization. Spatial downsampling between encoder stages is performed via 2D max-pooling, and decoder features are upsampled using bilinear interpolation. Skip connections transfer encoder feature maps to corresponding decoder stages.

\begin{figure}[ht]
    \centering
    \includegraphics[width=1\linewidth]{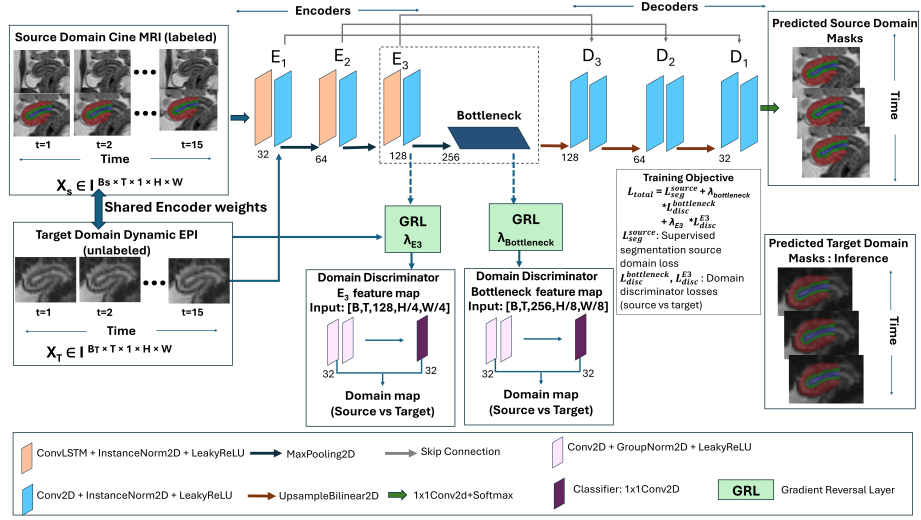}
    \caption{Architecture of the proposed U-Net–LSTM framework with multi-level adversarial domain discriminators enabling transfer from uterine layer segmentation from labeled dynamic cine MRI to unlabeled dynamic EPI.} 
    \label{architecture_fig}
\end{figure}

The adversarial domain discriminator blocks are placed at the third encoder stage\cite{zeiler2014visualizing} and the bottleneck, where feature representations are sufficiently abstract to encode structural, shape, and intensity characteristics of the uterine layers, while retaining temporal context through the C-LSTM. The labeled cine MRI data is the source, the unlabeled dynamic EPI data the target domain. Both domains are passed through the shared encoder weights, and feature maps at the third encoder stage and bottleneck are routed through gradient reversal layers (GRL) before being fed to their respective domain discriminators. During backpropagation, the GRL negates the gradient flowing back to the encoder, forcing the encoder to learn features that fool the discriminator into being unable to distinguish between cine MRI and dynamic EPI representations. Each discriminator block consists of two 2D convolutional layers followed by a $1\times1$ classifier convolution for binary domain prediction (source vs. target). Figure \ref{architecture_fig} illustrates the overall schematic of the proposed network and training objective. 

The training objective minimizes a segmentation loss on the labeled source domain, while simultaneously enforcing adversarial feature alignment across domains via binary cross-entropy loss. The segmentation loss is defined using a generalized weighted Dice \cite{sudre2017generalised}. The overall losses are:

\begin{equation}
    \mathcal{L}_{\text{total}} = \mathcal{L}_{\text{seg}}^{\text{source}} 
    + \lambda_{\text{bn}} \cdot \mathcal{L}_{\text{disc}}^{\text{bn}} 
    + \lambda_{\text{E3}} \cdot \mathcal{L}_{\text{disc}}^{\text{E3}}
\end{equation}
$\lambda_{\text{bn}}$ and $\lambda_{\text{E3}}$ are the domain loss weighting coefficients for bottleneck and third encoder stage discriminators, respectively, set to $\lambda_{\text{bn}} = 1.0$ and $\lambda_{\text{E3}} = 0.5$.

\noindent Segmentation loss is defined as:
\begin{equation}
\mathcal{L}_{\text{seg}}^{\text{source}} =
\mathcal{L}_{\text{GeneralizedDice}}
\left(
GT_{\text{mask}}^{\text{source}},
Pred_{\text{mask}}^{\text{source}}
\right)
\end{equation}
 $\text{GT}^{source}_{mask}$ is ground truth, $\text{Pred}^{source}_{mask}$ the predicted segmentation mask, respectively. Each domain discriminator loss is defined as:

\begin{equation}
    \mathcal{L}_{\text{disc}} = \frac{1}{2}\left(
    \mathcal{L}_{\text{BCE}}(D(\mathcal{F}_{source}), \mathbf{0}) + 
    \mathcal{L}_{\text{BCE}}(D(\mathcal{F}_{target}), \mathbf{1})
    \right),
\end{equation}

\noindent where $\mathcal{L}_{\text{BCE}}$ is the binary cross-entropy loss with $D(\mathcal{F}_{\text{source}})$ and $D(\mathcal{F}_{\text{target}})$ denoting the discriminator outputs for source and target domain feature maps.

\subsection{Data Preparation and Training Strategy}
\noindent For dynamic EPI data, the second echo was selected as a good compromise providing sufficient contrast for visualization of the three uterine layers: myometrium, junctional zone, and endometrium while maintaining sufficient signal-to-noise-ratio to facilitate segmentation. A mid-sagittal uterine slice was extracted from both cine and dynamic EPI sequences for subsequent analysis, both were pre-processed by intensity normalization and rescaling to the range of [0,1]. To match the spatial dimensions of the cine MRI ($256 \times 256$), the dynamic EPI images were further interpolated using bilinear interpolation. Additionally, qualitative visual inspection of the dynamic EPI data was performed to exclude cases severely affected by geometric distortions and bowel-gas susceptibility artifacts.

The source domain training set consisted of 99 cine MRI datasets acquired from 88 female participants, including 7 participants with repeated scans. For each case, the three uterine layers of the first 15 frames were manually annotated by one annotator and refined by a second. The target domain comprised 52 dynamic EPI cases (from 50 female participants) without segmentation masks for training, where similarly the first 15 frames of each case were used as  input. Seven-fold cross-validation was performed on the training dataset for 200 epochs with early stopping criteria. Given the limited size and variability of the source domain data, several data augmentation techniques were applied, including contrast-limited adaptive histogram equalization, horizontal and vertical flipping, and random gamma augmentation. Training was performed using the Adam optimizer with a learning rate of $1\times 10^{-4}$ and a batch size of 1 from scratch on an NVIDIA RTX 2080 Ti GPU. To stabilize adversarial training, a sigmoid-based ramp-up scheduling strategy, similar to the original domain adversarial network \cite{ganin2016domain}, was employed to progressively increase the influence of domain discrimination, allowing the network to first learn discriminative segmentation features of the uterine layers before enforcing domain alignment. 

\subsection{Segmentation Evaluation and T2* Analysis}
\begin{figure}[ht]
    \centering
    \includegraphics[width=1\linewidth]{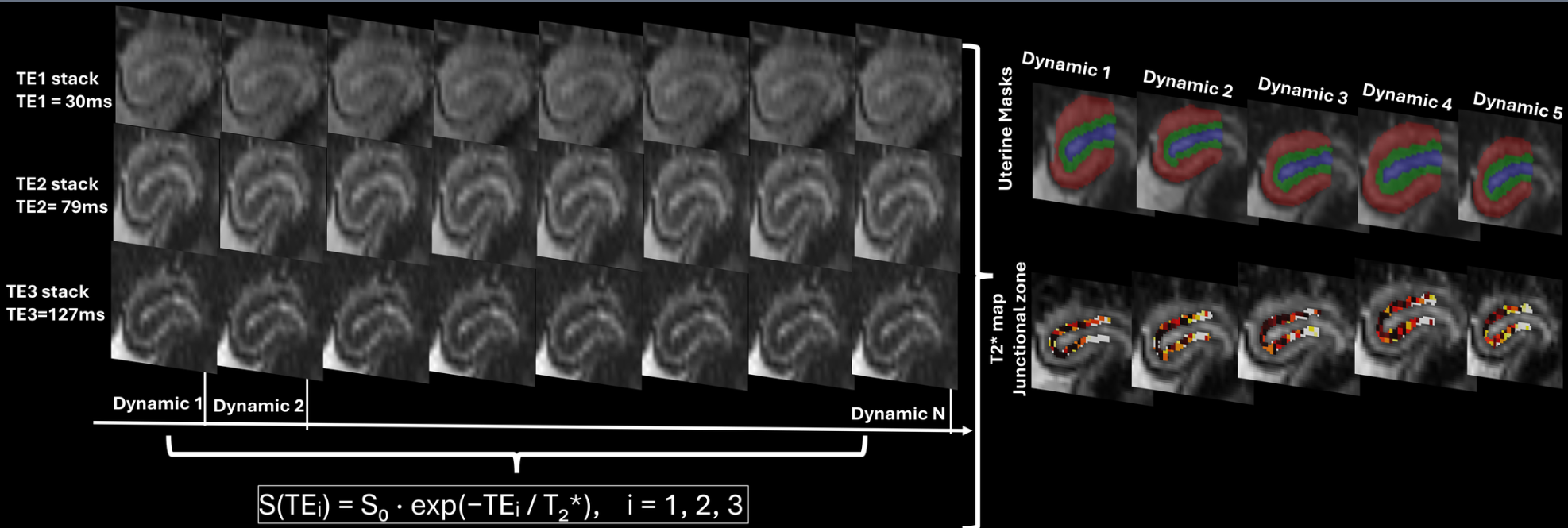}
    \caption{The uterine zones are shown on gradient-echo EPI images with three different echo times, along with uterine zone masks (red: myometrium, green: junctional zone, blue: endometrium) and the generated T2* maps across dynamics.}
    \label{fig:three_echo}
\end{figure}

Inference was performed using an ensemble of seven-fold trained networks, yielding predicted segmentation masks for the target domain dynamic EPI (n=52). The first 15 frames were manually segmented by the first annotator to generate ground-truth masks for evaluating network segmentation performance using the Dice similarity coefficient (DSC) and Jaccard index (JI). For dynamic T2* analysis, the predicted masks were post-processed by retaining the largest connected component and removing any small disconnected regions. T2* maps were computed for all 250 dynamics by mono-exponential decay fitting \cite{hutter2022dynamics}, as depicted in Figure \ref{fig:three_echo}. The segmented uterine layer masks obtained from the proposed network were applied to extract layer-specific T2* maps, and mean T2* values were computed per layer averaged across 250 dynamics for quantitative analysis. Dynamic analysis of T2* was performed on the junctional zone by computing the mean cross-sectional area over time and correlating it with mean  T2* values using Pearson correlation coefficient $r$, with statistical significance set at $p < 0.05$. Finally, the number of cases exhibiting positive, negative, and no significant correlation between junctional zone area and mean T2* was reported, along with the mean percentage change. 

\section{Results}
\subsection{Network Segmentation Analysis}
\begin{figure}[ht]
    \centering
    \includegraphics[width=1\linewidth]{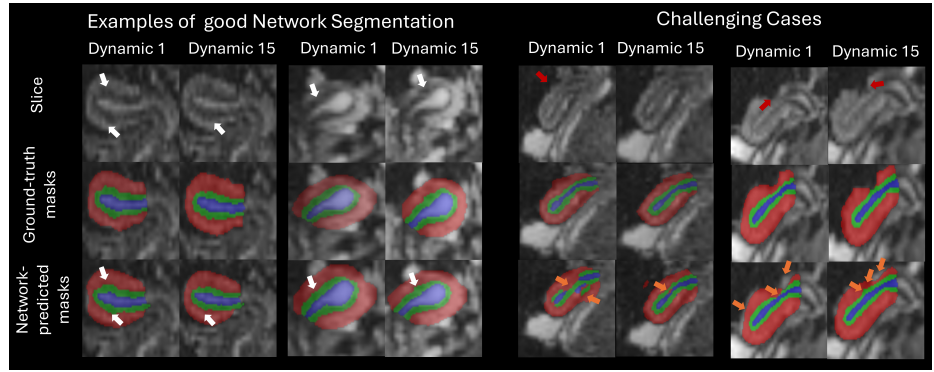}
    \caption{Ground truth and predicted masks over two dynamics for good and challenging cases. Red: myometrium, green: junctional zone, blue: endometrium. White arrows highlight accurately segmented regions of subtle uterine peristalsis. Red arrows indicate bowel-gas artifacts and orange arrows segmentation failure.}
    \label{fig:pred_masks}
\end{figure}

Table \ref{tab:seg_performance} summarizes the segmentation performance of the proposed network with and without domain discriminator at the bottleneck, and a baseline propagating the dynamic-1 mask. The proposed network achieved an overall DSC of $0.88 \pm 0.08$ and JI of $0.80 \pm 0.13$. The bottleneck domain discriminator ablation yielded an overall DSC of $0.85 \pm 0.26$ and JI of $0.80 \pm 0.29$, with  higher standard deviation. The mask-propagation baseline showed a reduced DSC ($0.76 \pm 0.10$) and JI ($0.63 \pm 0.13$) specifically for the junctional zone. The ablation study performed without domain discriminators demonstrated overall poor DSC ($0.47\pm0.36$) and JI ($0.38\pm0.34$) scores. Figure \ref{fig:pred_masks} illustrates a qualitative comparison for two distinct dynamics (t=1 and t=15), highlighting examples of successful and challenging segmentation. As observed in challenging cases, low inter-layer contrast, as well as overlying bowel gas artifacts, lead to misclassification of uterine layer regions and inconsistent delineation of the myometrium and junctional zone boundaries (indicated by arrows).

\begin{table}[ht]
\centering
\small
\setlength{\tabcolsep}{4pt}
\renewcommand{\arraystretch}{1.15}
\caption{Segmentation performance (mean$\pm$sd) of ablation studies on dynamic EPI (n=52) reported using Dice Similarity Coefficient (DSC) and Jaccard Index (JI).}
\label{tab:seg_performance}
\resizebox{\textwidth}{!}{%
\begin{tabular}{lcccccccc}
\hline
\multirow{2}{*}{Ablation} &
\multicolumn{2}{c}{Myometrium} &
\multicolumn{2}{c}{Junctional zone} &
\multicolumn{2}{c}{Endometrium} &
\multicolumn{2}{c}{Overall} \\
\cline{2-3}\cline{4-5}\cline{6-7}\cline{8-9}
 & DSC & JI & DSC & JI & DSC & JI & DSC & JI \\
\hline

\shortstack[l]{ Proposed \\(encoder-3 + bottleneck) }     & $0.87{\pm}0.08$ & $0.78{\pm}0.12$ & $0.86{\pm}0.09$ & $0.77{\pm}0.13$ & $0.90{\pm}0.08$ & $0.84{\pm}0.13$ & $0.88{\pm}0.08$ & $0.80{\pm}0.13$ \\
\shortstack[l]{Discriminator bottleneck} & $0.81{\pm}0.28$ & $0.75{\pm}0.29$ & $0.86{\pm}0.25$ & $0.82{\pm}0.28$ & $0.85{\pm}0.26$ & $0.84{\pm}0.29$ & $0.85{\pm}0.26$ & $0.80{\pm}0.29$ \\
Dynamic-1 mask propagation & $0.83{\pm}0.10$ & $0.72{\pm}0.13$ & $0.76{\pm}0.10$ & $0.63{\pm}0.13$ & $0.83{\pm}0.11$ & $0.72{\pm}0.15$ & $0.81{\pm}0.10$ & $0.69{\pm}0.14$ \\
No domain discriminators   & $0.43{\pm}0.30$ & $0.35{\pm}0.30$ & $0.49{\pm}0.36$ & $0.40{\pm}0.33$ & $0.47{\pm}0.36$ & $0.39{\pm}0.36$ & $0.47{\pm}0.36$ & $0.38{\pm}0.34$ \\
\hline
\end{tabular}%
}
\end{table}

\subsection{Dynamic T2* analysis}
\begin{figure}[ht]
    \centering
    \includegraphics[width=1\linewidth]{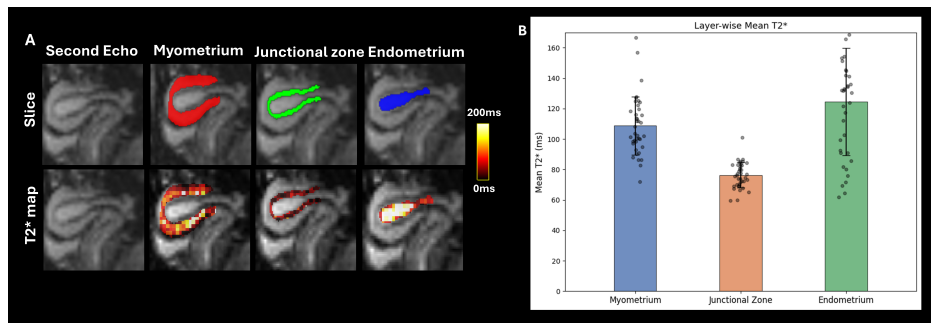}
   \caption{(A) Representative case showing uterine layer segmentation 
and corresponding layer-specific T2* maps. (B) Mean T2* per layer averaged across 250 dynamics for n=39 cases, with individual data points overlaid.}
    \label{fig:t2_map}
\end{figure}

\begin{figure}[ht]
    \centering
    \includegraphics[width=0.8\linewidth]{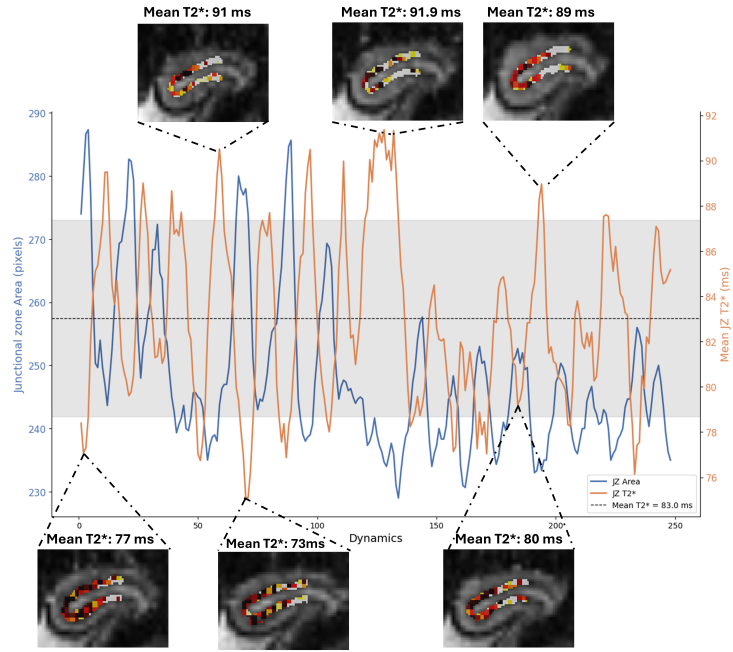}
    \caption{Exemplary case showing the inverse relationship between mean T2* and junctional zone area together with maps for the dynamics with distinct changes in T2*. Dotted line: mean T2*, shaded gray region: standard deviation over 250 dynamics.}
    \label{fig:dynamic_t2star}
\end{figure}

\noindent For dynamic T2* analysis, qualitative assessment identified 13 cases with bowel-gas artifacts during the mid-dynamics  (Figure \ref{fig:pred_masks}). These were excluded, resulting in a final cohort of n=39 cases for the reported T2* analysis. Figure \ref{fig:t2_map} shows an exemplary case with corresponding T2* maps of the myometrium, junctional zone, and endometrium, along with mean T2* values averaged across 250 dynamics. The overall mean T2* values were $108.69 \pm 19.20$ ms for the myometrium, $76.20 \pm 8.33$ ms for the junctional zone, and $124.00 \pm 35.29$ ms for the endometrium. Pearson correlation coefficient analysis between the junctional zone area and mean T2* for n=39 cases revealed a negative correlation in 14/39 (mean change in T2* values: $23.3\% \pm 6.7\%$, junctional zone area: $30.2\% \pm 18.4\%$), positive correlation in 7/39  (mean change in T2* values: $28.1\% \pm 25\%$, junctional zone area: $36.2\% \pm 15.8\%$) and no significant correlation in 18/39 (mean change in T2* values: $28.6\% \pm 23.6\%$, junctional zone area: $32.7\% \pm 22.6\%$). A negative correlation indicates that an increase in junctional zone area is accompanied by a decrease in mean T2* value, and vice versa. Fig.~\ref{fig:dynamic_t2star}illustrates a negative correlation between  T2*(percentage change: 28\%) and corresponding changes in the junctional zone area (percentage change: 29\%) across dynamics.


\section{Discussion and Conclusion}
\noindent An unsupervised adversarial domain adaptation framework for uterine layer segmentation in dynamic MRI, realized with a Unet-LSTM with multi-scale domain discriminators, outperformed both the bottleneck-only ablation and the dynamic-1 propagation baseline. The poor performance of the ablation study without domain discriminators confirmed the significant domain shift. The largest improvement in segmentation metrics over the propagation baseline was observed in the junctional zone, highlighting the importance of per-frame segmentation for subtle junctional zone dynamics. Furthermore, the multi-scale domain discriminator setup reduced the performance standard deviation, indicating more robust domain adaptation. Challenging cases remained where the network struggled to delineate boundaries between uterine layers due to low inter-layer contrast and with incomplete segmentation of the myometrium boundaries. These suggest the potential benefit of incorporating boundary-aware priors and loss functions \cite{sun2023boundary} to better constrain anatomically challenging regions, rather than relying predominantly on intensity-based features. 

The extracted layer-specific T2* maps were consistent with prior studies \cite{li2022cyclic,imaoka2012t2}. The comparatively lower T2* values in the junctional zone align with its known composition of densely packed fibers and reduced water content \cite{brown1991uterine}. The here proposed MEGE EPI acquisition enabled simultaneous evaluation of tissue T2* and uterine peristalsis. This inverse relationship between T2* and junctional zone are in a subset of cases may be attributed to wave-like contraction patterns within the junctional zone, supporting previous observations \cite{imaoka2012t2} that during active peristaltic activity, contractions transiently expel blood from the contracted region, reducing T2*, while the non-contracted state is associated with reduced junctional zone thickness and increased T2*. The variations in direction of observation may be influenced by the menstrual cycle phase and varying strength of uterine peristaltic activity, as reported in prior cine MRI studies \cite{de2023influence}. While pathological cases are already included into this methodological study, more stratified cohorts are required to better characterize the relationship between tissue composition and peristalsis across different pathological states. In conclusion, these results demonstrate the feasibility of combining functional information with uterine dynamics through automated domain-based unsupervised segmentation, to better understand uterine physiology.

\begin{credits}
\subsubsection{\ackname} This work was supported by the Endoki project, DFG Heisenberg [502024488], an ERC StG EARTHWORM [101165242], an ERC Proof-of-concept grant SYNCWORM [101293293] and CAIMed - Lower Saxony Center for Artificial Intelligence and Causal Methods in Medicine [ZN4257]. Code will be made publicly available upon publication. The authors have no competing interests to declare.
\end{credits}

%
%
%
\bibliographystyle{unsrt}
\bibliography{mybibliography}

\end{document}